\documentclass[showpacs,amssymb, amsmath,preprint]{revtex4-1}

\usepackage{amsmath}
\usepackage{amssymb}
\usepackage{epsfig}
\usepackage{subfigure}
\usepackage{graphicx}
\usepackage{color}
\usepackage{ulem}
\usepackage[titletoc]{appendix}

\begin{document}

\title{Enhancing noise-induced switching times in systems with distributed delays}
\author{Y.N. Kyrychko}
\email{y.kyrychko@sussex.ac.uk}
\affiliation{Department of Mathematics, University of Sussex, Falmer, Brighton, BN1 9QH, United Kingdom}
\author{I.B. Schwartz}
\affiliation{US Naval Research Laboratory, Code 6792, Nonlinear System Dynamics Section, Plasma Physics Division, Washington, DC 20375, USA}

\date{today}

\begin{abstract}
The paper addresses the problem of calculating the noise-induced switching
rates in systems with delay-distributed kernels and Gaussian noise. A general
variational formulation for the switching rate is derived for any distribution
kernel, and the obtained equations of motion and boundary conditions represent
the most probable, or optimal, path, which maximizes the probability of escape. Explicit
analytical results for the switching rates for small mean time delays are
obtained for the uniform and bi-modal (or two-peak) distributions. They
suggest that increasing the width of the distribution leads to an increase in
the switching times even for longer values of mean time delays for both
examples of the distribution kernel, and the increase is higher in the case of
the two-peak distribution. Analytical predictions are compared to the direct
numerical simulations, and show excellent agreement between theory and
numerical experiment.
\end{abstract}


\maketitle

\begin{quotation}
Many real world dynamical systems exhibit complex behavior often induced by intrinsic time delays. 
In addition, the majority of such systems are influenced by both external and internal random perturbations.
An important practical problem, therefore, is to understand how random disturbances
are organized such that the dynamics escape from a stable attractor and exhibit new behavior. Although the noise amplitudes are small, the
resulting change is a large fluctuation out of the  basin of attraction.
In this paper, we study the influence of noise-induced large fluctuations on dynamical systems, where the time delay is not taken as a constant but is rather chosen 
from a given distribution. We use a variational approach to calculate the switching rates out of the basin of attraction of the stable equilibrium for a general kernel of the delay distribution, as well as general nonlinearity. Taking two particular commonly-used examples of the distribution kernel, namely, 
the uniform and bi-modal kernels, we analyze how the width of the distribution affects the switching rates.
Our results  suggest that the switching is affected not only by the mean time
delay, but also by the width of the delay distribution.  Specifically, if the distribution width is increased, this leads to an increase in the switching times.
\end{quotation}

\section{Introduction}

Time-delayed models have been extensively applied in various disciplines,
including neuroscience, quantum optics, laser dynamics, mechanical and
chemical oscillators, population dynamics, mathematical epidemiology, and many
others \cite{enrneux}. In these systems, time delays are often used to account
for a finite propagation time, the time it takes for a signal to be processed
and fed back, and significant time delays can arise due to the separation
between the subsystems. In neural systems, delays represent the time it takes
for neural signals to propagate and be processed \cite{cam96, bootan1}, and in
semiconductor lasers the delay is determined by the propagation path, and is
much longer than the internal oscillation periods \cite{fischer1}. In
quantum networks, in order to model non-Markovian aspects of quantum dynamics,
one has to account for delayed interactions between network nodes when
exchanging photons, and time-delayed feedback control can be used to create
and stabilize entangled states \cite{zoller, knorr}. In order to describe
population fluctuations in ecological setting, it is important to include
maturation time, the time it takes to develop into a reproducing adult
\cite{wu14,wu2,kyrychko06}. In epidemiology, time delays can represent latency
or temporary immunity periods affecting the spread of the infection
\cite{peng17, kyrychko06,bauer}, and in gene regulatory networks they arise
during the transcription and translation of mRNA \cite{ott14, kiresh,
  levine16}. Finally, robotic systems such as swarms and their control have
attracting states that only can be captured with delayed communication and
control actuation~\cite{Klimka_2015,Hindes_Mothership}.      

In addition to intrinsic time delays, majority of real systems are also
affected by both external and internal random perturbations, which necessitate
the inclusion of noise into the dynamical models, and hence, stochastic
differential delay equations (DDEs)
are used to analyze a wide range of systems. For instance, gene regulatory
networks are often modeled as stochastic birth-death processes with time
delays \cite{ott14}, and anomalies during El Ni\~no have been analyzed using
delayed-oscillator models, excited by external weather noise
\cite{burgers99,chang, keane15}. Recently, time-delayed feedback control has
been applied to noise-induced chimera states in FitzHugh-Nagumo networks with
non-local coupling, and it has been shown that for some values of time delays,
it can lead to the so-called period-two coherence resonance chimera
\cite{anna17}. Much of the analysis of stochastic DDEs involving
local fluctuations has been done using 
delayed Fokker-Plank equations, and several small delay approximation methods
have been suggested. Furthermore, a perturbation theory method has been
developed to determine single time point probability densities and mean
delays, as well as autocorrelation functions in stochastic SDEs in the context
of delayed Fokker-Plank equations \cite{frank06,frank16}. 

In this paper we focus on understanding large fluctuations in dynamical systems
with distributed time delays, which have one or more stable steady states. In
the absence of noise, the solutions will go to a stable steady state; however,
when the noise is included, even if small in intensity, it leads to
fluctuations around the stable steady state. Furthermore, noise can force the
system out of the stable steady state, or it can lead to switching to another
stable steady state. This phenomenon is similar to the tipping process in dynamical systems, where relatively small changes in input can lead to sudden and disproportional changes in output, for example, due to a slow variation in parameter (B-tipping) or inclusion of noise (N-tipping) \cite{ashwin,ott2014,sieber2016}.  For systems not in thermal equilibrium, as in the case
of noise-induced switching between states, 
the probability distribution is no longer of the Boltzmann form. Many results
have been  obtained for dynamical systems without delay driven by white
Gaussian noise and for Markovian reaction and population systems,
cf. \cite{Freidlin_book,Wentzell1976,Graham1984a,Graham1989a,Dykman1980,Dykman1994c,Dykman2008,Jauslin1987,Maier1993a,Maier1997,Touchette2009,Kamenev2011}. 

On the other hand, noise-induced switching has been studied for  bistable
systems in the presence of time delay to demonstrate
residence times depending on the noise level, as well as to analytically
calculate the expressions for the autocorrelation function and power spectrum
\cite{pik01, mass03}. Switching entire behaviors for complex discrete
delay-coupled swarming systems has been observed in~\cite{lindley2013noise}. Noise-induced switching rates between stable steady
states, and the rates of noise-induced extinction in the case of constant time
delays, have recently been analyzed in general in \cite{ira15}. In this paper
we will study a time-delayed system driven by noise, where the time delay is
given by an integral with a memory kernel in the form of the prescribed delay
distribution function. The research in this paper  is motivated by the
  effects of distributed delays on the system dynamics in various settings; for example, in models coupled oscillators \cite{kyr11,kyr13,kyr14}, in traffic dynamics to describe the memory effects on drivers \cite{sipahi}, to represent long-range interactions between neurons \cite{bootan1, sueann}, modeling waiting times in epidemiological models \cite{blyuss2010}, as well as maturation periods in population dynamics modeling \cite{faria, gourley}. 

The outline of the paper is as follows. In Section~\ref{sec2} we apply variational approach to formulate the problem in order to calculate the rate of switching out of the basin of attraction of the steady state for a general delay distribution and general nonlinearity. In Section~\ref{sec3}, we consider the case of the uniformly distributed delay kernel, and show how the width of the distribution influences the switching rate. The case of the two-peak delay distribution is analyzed in Section~\ref{sec4}, and the results are compared to the case of the uniform distribution. For both types of distributions, we perform numerical simulations, which are presented in Section~\ref{sec5}, and the agreement between theory and analytical results is excellent. 

\section{General case of the delay distribution kernel}\label{sec2}

We consider a switching process in a one-dimensional system with distributed time delay and noise, which has the form
\begin{equation}\label{eq1}
\dot{x}(t)=f\left(x(t),\int_{0}^{\infty}g(\sigma)x(t-\sigma)d\sigma\right)+\eta(t),
\end{equation}
where the distribution kernel $g(\cdot)$ is assumed to be non-negative and normalized to unity, i.e.
\[
g(s)\geq 0, \quad\quad \int_{0}^{\infty} g(s)ds =1,
\]
$\eta(t)$ is the noise, and $f(\cdot)$ is a nonlinear function. If
$g(u)=\delta(u)$, one obtains a system without time delays, and if
$g(u)=\delta(u-\tau)$, this leads to the case of discrete time delay. 

In the
case of Gaussian noise, if $D$ is the noise intensity, one can characterize
noise realizations $\eta(t)$ by its probability density functional
$\mathcal{P}_{\eta}[\eta(t)] \approx \exp(-\mathcal{R}_{\eta}/D)$ with
\begin{equation}
\mathcal{R}_{\eta}[\eta(t)]=\frac{1}{4}\int_{-\infty}^{\infty} dt dt'\eta(t)\hat{\mathcal{F}}(t-t')\eta(t'),
 \label{eq:GN}
\end{equation}
where $\hat{\mathcal{F}}(t-t')/2D$ is the inverse of the pair correlator of
$\eta(t)$~\cite{feynman}. Noise intensity $D$ is assumed to be sufficiently
small, so that sample paths will limit on an optimal path as $D\to 0$. 

In the absence of noise, we consider a situation in which the system
possesses a stable steady state, $x_{A}$, and a saddle point,
$x_{S}$, satisfying $f(x_{A},x_{A})=f(x_{S},x_{S})\equiv
0$. For the topological switching setup, we suppose the saddle lies on the
boundary of the basin of attraction of $x_{A}$. That is, we will consider
starting initially in the basin of attraction of attractor $x_A$, and
consider the dynamics of switching as escape from the basin of attraction, which we define as a
large fluctuation. In the asymptotic limit
as the noise intensity goes to zero, the path away from $x_A$ will pass
through the saddle point, $x_S$.

When compared to the effective barrier height, we assume the  noise is small, and  
the mean time to switch  is much longer than the
relaxation time. Thus, the occurrence of switching is expected to be a rare
event. Assuming the event lies in the tail of the distribution, the probability  of a switching is an exponential
distribution, given as
\begin{equation}\label{rate1}
\mathcal{P}_{x}[x] \propto \exp(-R/D), \quad\quad R=\min \mathcal{R}[x,\eta,\lambda],
\end{equation}
where
\begin{equation}
\mathcal{R}[x,\eta,\lambda]=\mathcal{R}_{\eta}[\eta(t)] +\int^{\infty}_{-\infty} \lambda(t)\left[\dot{x}(t)-f\left(x,\int_{0}^{\infty}g(\sigma)x(t-\sigma)d\sigma\right)-\eta(t)\right]dt,
\label{eq:GenR}
\end{equation}
where $\lambda(t)$ is a time-dependent Lagrange multiplier. 
Note that the noise source $\eta(t)$ may be from any continuous or
  discrete distribution. Here we assume uncorrelated Gaussian noise defined in
equation~(\ref{eq:GN}) so that we substitute it into equation~(\ref{eq:GenR}) as
\begin{equation}
\mathcal{R}_{\eta}[\eta(t)] = \frac{1}{4}\int\eta^2(t)dt.
\label{eq:UGN}
\end{equation}
The main goal to determine the probability of switching is to compute the
  exponent $R$, which, for reasons that will be made clear below, we define
  as the action, similar to the action in classical mechanics.

\subsection{The Variational Equations of Motion}

To simplify the analysis, we define the following convolution:
\begin{equation}
(g \circledast x)(t) \equiv  \int_{0}^{\infty}g(\sigma)x(t-\sigma)d\sigma .
\label{eq:conv}
\end{equation}
In order to find the exponent $\mathcal{R}$ in equations~(\ref{eq:GenR}) and
(\ref{eq:UGN}), we look for equations that describe the maximum probability of
reaching the state $x_{S}$ from the initial state $x_{A}$. 
The variation $(\delta\mathcal{R})$ is obtained by varying deviations from the
path that minimize $\mathcal{R}$. Variation with respect to noise $\eta(t)$ is $\eta(t)=2\lambda(t)$.

The variation with respect to the Lagrange multiplier $\lambda(t)$  has the form
\[
\dot{x}(t)=f\left(x(t),(g \circledast x)(t)\right)+2\lambda(t).
\]
Finally, we look at the variation with respect to $x$, which is given by
\begin{align*}
\frac{\delta\mathcal{R}}{\delta x} &= \mathcal{R}[x+\mu,\eta,\lambda]-\mathcal{R}[x,\eta,\lambda]\\
&= \int_{-\infty}^{\infty}\lambda(t) \left[\dot{\mu}(t) -f\left(x(t)+\mu(t),(g \circledast x)(t)+(g \circledast \mu)(t)\right)\right.\\
&\left. + f\left(x(t),(g \circledast x)(t)\right)\right] dt.
\end{align*}
Taylor expanding the last expression gives
\begin{align}\label{eq33}
\frac{\delta\mathcal{R}}{\delta x} &=\int_{-\infty}^{\infty} \lambda(t)\int_{0}^\infty g(\sigma)\left[\dot{\mu}(t)-\mu(t)f_{1}\left(x(t),(g \circledast x)(t)\right)\right.\\\nonumber
&\left. -f_{2}\left(x(t),(g \circledast x)(t)\right)\mu(t-\sigma)\right]d\sigma dt +\mathcal{O}(\mu^2),
\end{align}
where $f_{i}(\cdot,\cdot)$, $i=1,2$ are 
\[
f_{1}=\frac{\partial f\left(x(t),(g \circledast x)(t)\right)}{\partial x} \quad\text{and}\quad
f_{2}=\frac{\partial f\left(x(t),(g \circledast x)(t)\right)}{\partial (g \circledast x)(t)}.
\]
Evaluating the second integral in (\ref{eq33}) gives the equation of motion for $\lambda$ in the form:
\begin{align}\label{eq44}
-\dot{\lambda}(t)&=\lambda(t) f_{1}\left(x(t),(g \circledast x)(t)\right)\\ \nonumber&+
\int_{0}^{\infty}\lambda(t+\sigma)g(\sigma)f_{2}\left[x(t+\sigma),(g \circledast x)(t+\sigma)\right]d\sigma.
\end{align}
Note that unlike the equations for the noise and the state functions,  the last equation
contains both delayed and advanced terms, making it an acausal equation of
motion for $\lambda(t)$. Combining the derived variational results leads to a system of equations in the form

\begin{align}
& \dot{x}(t)=f\left(x,(g \circledast x)(t)\right)+2\lambda(t),\\ \label{eq:GEOM1}
&\dot{\lambda}(t)=-\lambda (t) f_{1}\left(x(t),(g \circledast x)(t)\right)\\ \label{eq:GEOM2} &-
\int_{0}^{\infty}\lambda(t+\sigma)g(\sigma)f_{2}\left[x(t+\sigma),(g
  \circledast x)(t+\sigma)\right]d\sigma, \nonumber
\end{align}

which has the Hamiltonian
\begin{equation}
H\left(x, (g \circledast x)(t), \lambda\right)=\lambda^2+\lambda f\left(x(t),(g \circledast x)(t)\right).
\label{eq:GH}
\end{equation}

\subsection{Boundary conditions}

To solve for the most likely, or optimal, path along which the system switches from the attractor,
 we assume we start at or near $x_A$, and compute the path to the 
saddle point $x_S$ using the equation (\ref{eq:GEOM1}). In order to compute the actual path, boundary conditions are required on the
real line. We assume that at steady state, the noise goes to zero. Therefore,
we have that as $t \rightarrow \infty$, $(x(t),\lambda(t)) \rightarrow (x_S,0)$. On
the other hand, as $t \rightarrow -\infty$, we have $(x(t),\lambda(t))
\rightarrow (x_{A},0))$.

When noise $\eta(t) \equiv 0$, we can linearize the equation~(\ref{eq1}) about a steady state, $\bar{x}$:
\begin{equation}
\dot{X}(t) = f_{1}(\bar{x},\bar{x})X(t) + f_{2}(\bar{x},\bar{x})(g \circledast X)(t).
\label{eq:LinEq}
\end{equation}
In particular, when $\bar{x}=x_A$, the spectrum of the characteristic equation
has values in the left half of the complex plane. In contrast, when $\bar{x}=x_S$, at
least one eigenvalue has positive real part. 

On the other hand, since the equation~(\ref{eq44}) is linear in $\lambda$, at the
steady state $\bar{x}$ we have
\begin{equation}
\dot{\Lambda}(t)=-\Lambda(t)f_1(\bar{x},\bar{x}) - f_2(\bar{x},\bar{x})\int_{0}^{\infty}g(\sigma)\Lambda(t+\sigma)d\sigma.
\end{equation}
Assuming $\Lambda(t)=e^{\alpha t}$, the characteristic equation for $\Lambda$ is,
\begin{equation}\label{char1}
-\alpha-f_{1}(\bar{x},\bar{x}) -f_{2}(\bar{x},\bar{x})[\{\mathcal{L}g\}(-\alpha)]=0,
\end{equation}
where $\alpha$ is an eigenvalue of a Jacobian, and 
\[
\{\mathcal{L}g\}(z)=\int_{0}^{\infty}e^{-zu}g(u)du,
\]
is the Laplace transform of the function $g(u)$. Moreover, it is easy to show
that the spectrum of $\Lambda$ have opposite signs to that of $X$. Therefore,
the attractor becomes a saddle, and saddle remains a saddle. This result will
hold true if the mean value of the delay distribution is small.

The interpretation of the local  saddle structure and the escape path is such that at attractor $x_A$,
the conjugate variable supplies an unstable direction for escape through
$x_S$. Escape occurs along the most probable path which connects saddles
$(x_A,0)$ to $(x_S,0)$ as a heteroclinic orbit. That is, as $t \rightarrow
-\infty$, the path approaches $(x_A,0)$ along its unstable manifold, while as
$t \rightarrow \infty$, the path asymptotes to $(x_S,0)$ along the stable
manifold. 

\subsection{Perturbation Theory}

Taking into account the above-mentioned assumptions on the equations of motion
and asymptotic boundary conditions, the problem of finding the action
$\mathcal{R}$ is formulated as calculating the solutions to a nonlinear
two-point boundary problem. We assume that this solution exists in the
non-delayed case, and when the delay is non-zero, the corresponding solution
remains close to the non-delayed solution. Thus the variational problem of
finding the action is formulated as finding its perturbation to non-delayed case. 
We assume that if $x(t)$ is a solution for a non-delayed problem, then the perturbations  $\displaystyle{\left[x(t)-\int_{0}^{\infty}g(z)x(t-z)dz\right]}$ should remain small. The action can be written as the following perturbation problem
\[
\mathcal{R}[x,\eta,\lambda]=\mathcal{R}_{0}[x,\eta,\lambda]+\mathcal{R}_{1}[x,\eta,\lambda],
\]
where
\begin{equation}
\mathcal{R}_{0}[x,\eta,\lambda]=\frac{1}{4}\int\eta^2(t)
dt+\int\lambda(t)[\dot{x}(t)-f(x,x)-\eta(t)]dt,\label{eq:R0}
\end{equation}
and
\begin{equation}
\mathcal{R}_{1}[x,\eta,\lambda]=\int\lambda(t)\left[f(x(t),x(t))-f\left(x(t),(g \circledast x)(t)\right)\right]dt.\label{eq:R1}
\end{equation}
The minimizing solution can be found by first analyzing the equations that
minimize $\mathcal{R}_{0}$, which are denoted by
$[x_{o},\eta_{o},\lambda_{o}]$, and then evaluating the first order correction
at the zeroth order solution; i.e., $\mathcal{R}_{1}[x_0,\eta_0,\lambda_0]$.

From equations~(\ref{eq:R0}) and (\ref{eq:GH}), it is easy to see that the the optimal path  for the non-delayed
case is given by $\lambda_0 = -f(x_0,x_0)$, implying that $\dot{x}_0 =
-f(x_{0},x_{0})$, which is just a time-reversed trajectory from the zero delay
case. The optimal noise is then given by $\eta_{0}(t)=2 \lambda_0 (t)$.

In order to make further analytical progress, we need to specify a particular choice of the distribution kernel $g(u)$ and of the function $f(\cdot)$. In this paper we consider two particular distributions, namely, a uniform and a two-peak distribution kernel, and a quadratic nonlinearity in the function $f(\cdot)$.

\section{Uniformly distributed delay}\label{sec3}

In this section we consider the case of the uniformly distributed kernel that can be written as follows,
\begin{equation}\label{uniform}
g(u)=\left\{
\begin{array}{l}
\displaystyle{\frac{1}{2\rho}, \quad \tau-\rho\leq u\leq \tau+\rho,}\\\\
0,\quad\text{otherwise.}\end{array}
\right.
\end{equation}

With this distribution kernel, we choose the function $f(\cdot)$ to be 
\begin{equation}\label{non1}
f\left(x(t),(g \circledast x)(t)\right)=x(1-x)-\frac{\gamma}{2\rho}\int_{\tau-\rho}^{\tau+\rho}x(t-z)dz.
\end{equation}
Substituting the expression (\ref{non1}) into the right-hand side of the equation (\ref{eq1}), one obtains
\begin{equation}\label{eq222}
\dot{x}(t)=x(1-x)-\frac{\gamma}{2\rho}\int_{\tau-\rho}^{\tau+\rho}x(t-z)dz+\eta(t).
\end{equation}

\subsection{Steady State Stability}

In the absence of noise, the equation (\ref{eq222}) has two steady states, 
\[
x_{A}=1-\gamma\quad\text{and}\quad x_{S}=0,\quad 0\leq\gamma\leq 1.
\]
To study the stability of these steady states, we substitute the Laplace transform of the distribution kernel
\[
\{\mathcal{L}g\}(-\alpha)=\frac{1}{2\rho\alpha}e^{\alpha\tau}(e^{\alpha\rho}-e^{-\alpha\rho})=e^{\alpha\tau}\frac{\sinh(\alpha\rho)}{\alpha\rho}.
\]
into the characteristic equation (\ref{char1}) for the conjugate variable,
$\Lambda$. (Note that since the spectra for $X$ and $\Lambda$ characteristic
equations are of opposite sign, we only examine the  one for $\Lambda$.) Using the Taylor expansion for small mean time delays (and hence, for small $\rho$), we can simplify the characteristic equation and find eigenvalues $\alpha$ as
\[
-\alpha -1+2\bar{x}+\gamma(1+\alpha\tau)=0 \quad\Longrightarrow\quad \alpha=\frac{1-2\bar{x}-\gamma}{\gamma\tau-1}.
\]

At the steady state $\bar{x}=x_{A}$, we have
$\alpha=\frac{1-\gamma}{1-\gamma\tau}>0$ for $\Lambda$, and
$\alpha=\frac{1-\gamma}{1-\gamma\tau}<0$ for $X$.  
Similar but opposite sign results hold for $x_S$.

 Therefore, when $t\to-\infty$, the solution of the linearized problem tends to $(x_A,0)$, and the solution of the linearized problem tends to $(x_{S},0)$ as $t\to\infty$.

\subsection{Perturbation of the Action}
In order to find the optimal path from $x_{A}$ to $x_{S}$, which minimizes
$\mathcal{R}_{0}$, we use the fact that the optimal path solutions satisfy $\dot{x}_{o}(t)=-f(x_{o}(t),x_{o}(t))$ and $\lambda_{o}=-f(x_{o}(t),x_{o}(t))$. 

For the vector field without delay, we find that 
\begin{equation}\label{xo}
x_{o}=\frac{x_{A}}{1+e^{x_{A}t}},
\end{equation}
that as $t\to\infty$, $x_{o}(t)\to 0$, and as $t\to-\infty$, $x_{o}(t)\to x_{A}$
and
\begin{equation}
\mathcal{R}_{0}(\tau,\rho)=\frac{x_{A}^{3}}{3}.
\label{R0Ex1}
\end{equation}
Since
\[
f(x(t),x(t))-f\left(x(t),(g \circledast x)(t)\right)=-\gamma\left[x(t)-\frac{1}{2\rho}\int_{\tau-\rho}^{\tau+\rho}x(t-z)dz\right],
\]
the first order action can be found as
\begin{align}
\mathcal{R}_{1}(\tau,\rho)&=-2\gamma\int_{-\infty}^{\infty}
\dot{x}_{o}(t)\left[x_{o}(t)-\frac{1}{2\rho}\int_{\tau-\rho}^{\tau+\rho}x_{o}(t-z)dz\right]dt.\\ \nonumber
&= -2\gamma\int_{-\infty}^{\infty} -x_{o}(x_{A}-x_{o})\left[x_{o}(t)
- x_{A}-\frac{1}{2\rho}\ln\left(\frac{1+e^{x_{A}(t-\tau-\rho)}}{1+e^{x_{A}(t-\tau+\rho)}}\right)\right]dt.\label{eq225}
\end{align}

Combining the above calculations, the expression for $\mathcal{R}(\tau,\rho)$ becomes
\begin{equation}
\mathcal{R}(\tau,\rho)=\frac{x_{A}^{3}}{3}-\gamma
x_{A}^{2}
-\frac{\gamma}{\rho}\cdot\frac{x_{A}^{2}(\tau+\rho)}{e^{x_{A}(\tau+\rho)}-1}+\frac{\gamma}{\rho}\cdot\frac{x_{A}^{2}(\tau-\rho)}{e^{x_{A}(\tau-\rho)}-1},
\end{equation}
where we have an explicit dependence on the mean time delay $\tau$ and the distribution width $\rho$. For small distribution widths $\rho$, this expression can be further expanded as
\begin{align}\label{approx22}
\mathcal{R}(\tau,\rho)&\approx  \frac{x_{A}^{3}}{3}-2\gamma \left(-\frac{x_{A}^{2}}{2}+\frac{x_{A}^{2}e^{x_{A}\tau}(e^{x_{A}\tau}-x_{A}\tau-1)}{(e^{x_{A}\tau}-1)^2}\right)\nonumber \\\\\nonumber
&+ \frac{1}{3}\frac{x_{A}^{4}e^{x_{A}\tau}(-3e^{2x_{A}\tau}+3+x_{A}\tau e^{2x_{A}\tau}+4x_{A}\tau e^{x_{A}\tau}+x_{A}\tau)}{(e^{x_{A}\tau}-1)^4}\rho^2+\mathcal{O}(\rho^4).
\end{align}
If $\rho=0$, then we recover the same expression as in \cite{ira15}, which was derived for the case of the discrete time delay. 

\section{A two-peak delay distribution kernel}\label{sec4}

In this section we analyze the effect of distributed delay on the switching rates by considering a distribution kernel $g(u)$ in the form of two discrete time delays with an average delay $\tau$, which are separated by a time interval $2\rho$~\cite{fahse}.
Under this approximation, the distribution kernel $g(u)$ can be written as
\begin{align*}
g(u)=\frac{\delta(u-\tau-\rho)+\delta(u-\tau+\rho)}{2},
\end{align*}
where $\delta$ is the Dirac delta function.
For simplicity, we introduce the notation as follows
\[
\tau+\rho=\tau_{1}, \quad \tau-\rho=\tau_{2}, 
\]
and
\[
x_{\tau_{1}}=x(t-\tau_{1}), \quad x_{\tau_{2}}=x(t-\tau_{2}).
\]
The calculations for the variation with respect to noise and $\lambda(t)$ will
be similar to those in Section~\ref{sec2}, so we just state the one variation with
respect to $x(t)$, since the problem now involves multiple delays. Looking at the deviation with respect to $x(t)$ gives:
\begin{align*}
\frac{\delta\mathcal{R}}{\delta x} &= -\int\mu(t)\left[
\dot{\lambda}(t)+\lambda(t)\frac{\delta f(x,x(t-\tau_{1}),x(t-\tau_{2}))}{\delta x(t)}\right.\\
&\left.-\lambda(t+\tau_{1})\frac{\delta f(x(t+\tau_{1}),x(t),x(t+2\rho))}{\delta x(t-\tau_{1})}\right.\\
&\left. -\lambda(t+\tau_{2})\frac{\delta f(x(t+\tau_{2}),x(t-2\rho),x(t))}{\delta x(t-\tau_{2})}\right]dt.
\end{align*}
Now, the equations of motion can be written in the following way
\begin{align*}
&\dot{x}(t)= f(x(t),x(t-\tau_{1}),x(t-\tau_{2}))+2\lambda(t),\\\\
& \dot{\lambda}(t)=-\lambda(t)\frac{\delta f(x,x(t-\tau_{1}),x(t-\tau_{2}))}{\delta x(t)}\\
&-\lambda(t+\tau_{1})\frac{\delta f(x(t+\tau_{1}),x(t),x(t+2\rho))}{\delta x(t-\tau_{1})}
-\lambda(t+\tau_{2})\frac{\delta f(x(t+\tau_{2}),x(t-2\rho),x(t))}{\delta x(t-\tau_{2})}.
\end{align*}
We remark that since the problem is one with multiple delays, the equations of
motion can be derived using the Hamiltonian:
\[
H(x,x_{\tau_{1}},x_{\tau_{2}},\lambda)=\lambda^2+\lambda f(x,x_{\tau_{1}},x_{\tau_{2}}),
\]
with the general acausal equations of motion: 
\begin{align*}
& \dot{x}_{o}=\frac{\partial H}{\partial \lambda} (x_{o}, x_{o_{\tau_{1}}},x_{o_{\tau_2}},\lambda_{o}),\\
&\dot{\lambda}_{o}= -\frac{\partial H}{\partial x} (x_{o}, x_{o_{\tau_{1}}},x_{o_{\tau_2}},\lambda_{o})-\frac{\partial H}{\partial x_{\tau_{1}}}(x_{o}(t+\tau_{1}),x_{o}(t),x_{o}(t+2\rho),\lambda_{o}(t+\tau_{1}))\\
&-\frac{\partial H}{\partial x_{\tau_{2}}}(x_{o}(t+\tau_{2}),x_{o}(t-2\rho),x_{o}(t),\lambda_{o}(t+\tau_{2})).
\end{align*}

In order to compute the optimal path, we assume that the solution exists for
$\tau_{1,2}=0$, and the solutions for $\tau_{1,2}\neq 0$ remain close to the
non-delayed solution, and as before we assume that perturbations to the optimal
path when $\tau_{1,2}\neq 0$ remain small. This means that if $x(t)$ is a
solution for $\tau_{1,2}=0$, then the perturbations $\delta x_{\tau_{i}}\equiv
x(t)-x(t-\tau_{i})$, $i=1,2$, should remain small. The action is  formulated using the
perturbation problem given in equations~(\ref{eq:R0}) and (\ref{eq:R1}) for
$\mathcal{R}_{0}[x,\eta,\lambda]$ and $\mathcal{R}_{1}[x,\eta,\lambda]$,
where
\[
\mathcal{R}_{0}[x,\eta,\lambda]=\frac{1}{4}\int\eta^2(t) dt+\int\lambda(t)[\dot{x}(t)-f(x,x,x)-\eta(t)]dt,
\]
and
\[
\mathcal{R}_{1}[x,\eta,\lambda]=\int\lambda(t)[f(x(t),x(t),x(t))-f(x(t),x(t-\tau_{1}),x(t-\tau_{2}))]dt,
\]
so the action now explicitly depends on two constant time delays $\tau_{1}$ and $\tau_{2}$.
In order to compare the results of the two-peak distribution kernel with the results on uniform distribution obtained in Section~\ref{sec3}, as well as to the case of the single discrete time delay considered in \cite{ira15}, we consider function $f$ as follows,
\[
f(x,x_{\tau_{1}},x_{\tau_{2}})=x(1-x)-\frac{\gamma}{2}(x_{\tau_{1}}+x_{\tau_{2}}).
\]
When $\eta(t)=0$,  there are  two steady states
\[
x_{A}=1-\gamma\quad\text{and}\quad x_{S}=0,\quad 0\leq\gamma\leq 1.
\]
Linearizing near the steady states, using the characteristic equation (\ref{char1}), yields
\[
-\alpha -1+2\bar{x}+\frac{\gamma}{2}(1+\alpha\tau_{1}+1+\alpha\tau_{2})=0 \quad\Longrightarrow\quad \alpha=\frac{1-2\bar{x}-\gamma}{(\gamma/2)(\tau_{1}+\tau_{2})-1}=\frac{1-2\bar{x}-\gamma}{\gamma\tau-1},
\]
and the stability characteristics of the attractor and saddle are exactly the
same as in the previous section.

In the absence of delay, the computation of $\mathcal{R}_{0}$ is the same as
before, and we only need to compute $\mathcal{R}_{1}$.  Since
\[
f(x,x,x)-f(x,x_{\tau_{1}},x_{\tau_2})=-\frac{\gamma}{2}[(x-x_{\tau_{1}})+(x-x_{\tau_{2}})],
\]
 we can find the first order action as
\begin{align*}
\mathcal{R}_{1}(\tau_{1},\tau_{2})&=-\gamma \int_{-\infty}^{\infty}
\dot{x}_{o}(t)[x_{o}(t)-x_{o}(t-\tau_{1})]dt -\gamma \int_{-\infty}^{\infty}
\dot{x}_{o}(t)[x_{o}(t)-x_{o}(t-\tau_{2})]dt.\\
&=2\gamma \int_{-\infty}^{\infty} \frac{x_{A}^{3}}{(1+e^{x_{A}t})^3}dt -\gamma x_{A}^{3}\int_{-\infty}^{\infty} \frac{e^{x_{A}t}}{(1+e^{x_{A}t})^2(1+e^{x_{A}(t-\tau_{1})})}dt\\
&-\gamma x_{A}^{3}\int_{-\infty}^{\infty} \frac{e^{x_{A}t}}{(1+e^{x_{A}t})^2(1+e^{x_{A}(t-\tau_{2})})}dt.
\end{align*}
Evaluating the integrals in the last expression gives the first order action $\mathcal{R}_{1}(\tau_{1},\tau_{2})$
and the full action to the first order in $\delta x(t)$ can be found as
\begin{equation}\label{approx}
\mathcal{R}(\tau_{1},\tau_{2})\approx\frac{x_{A}^{3}}{3}-\gamma x_{A}^{2}\left(-1+\frac{e^{x_{A}\tau_{1}}[-x_{A}\tau_{1}-1+e^{x_{A}\tau_{1}}]}{(e^{x_{A}\tau_{1}}-1)^2} + \frac{e^{x_{A}\tau_{2}}[-x_{A}\tau_{2}-1+e^{x_{A}\tau_{2}}]}{(e^{x_{A}\tau_{2}}-1)^2}\right),
\end{equation}

\noindent where $\tau_{1}=\tau+\rho$ and $\tau_{2}=\tau-\rho$. If the width $\rho=0$, the time delays $\tau_{1}$ and $\tau_{2}$ coincide, and the expression (\ref{approx}) becomes the same as the one derived in \cite{ira15} for the case of the single time delay.

\section{Numerical simulations}\label{sec5}

\begin{figure}[ht]
\includegraphics[scale=0.3]{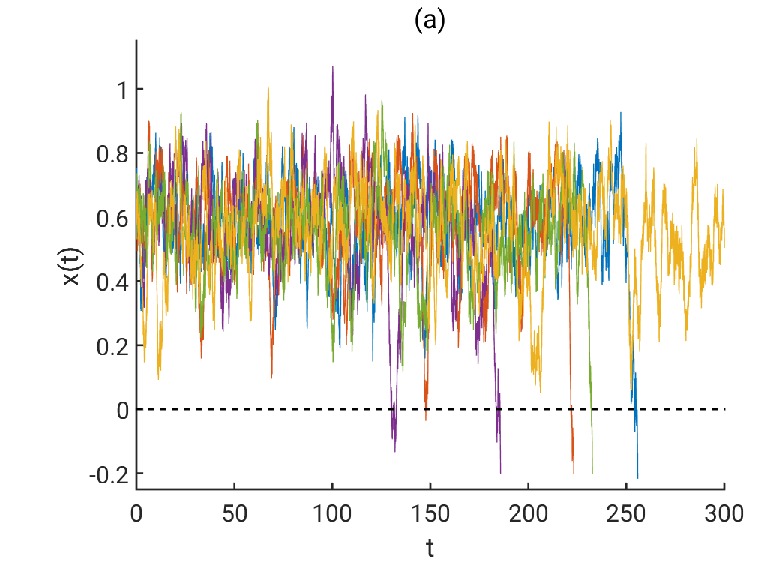}  \includegraphics[scale=0.6]{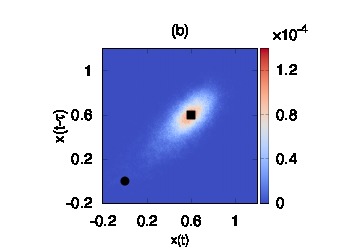} 
\caption{(a) Sample switching time series in (\ref{eq222}) and (b) a
  probability density of switching obtained by numerical simulations for the uniformly distributed kernel. Black squares (circles) denote the
  attractor (saddle point). Parameters are: $\gamma=0.4$, $\sqrt{2D}=0.12$, $\tau=0.8$, $\rho=0.5$.} 
\label{fig:TS_Uniform}
\end{figure}

\begin{figure}[ht]
\hspace{1cm}\includegraphics[width=140mm]{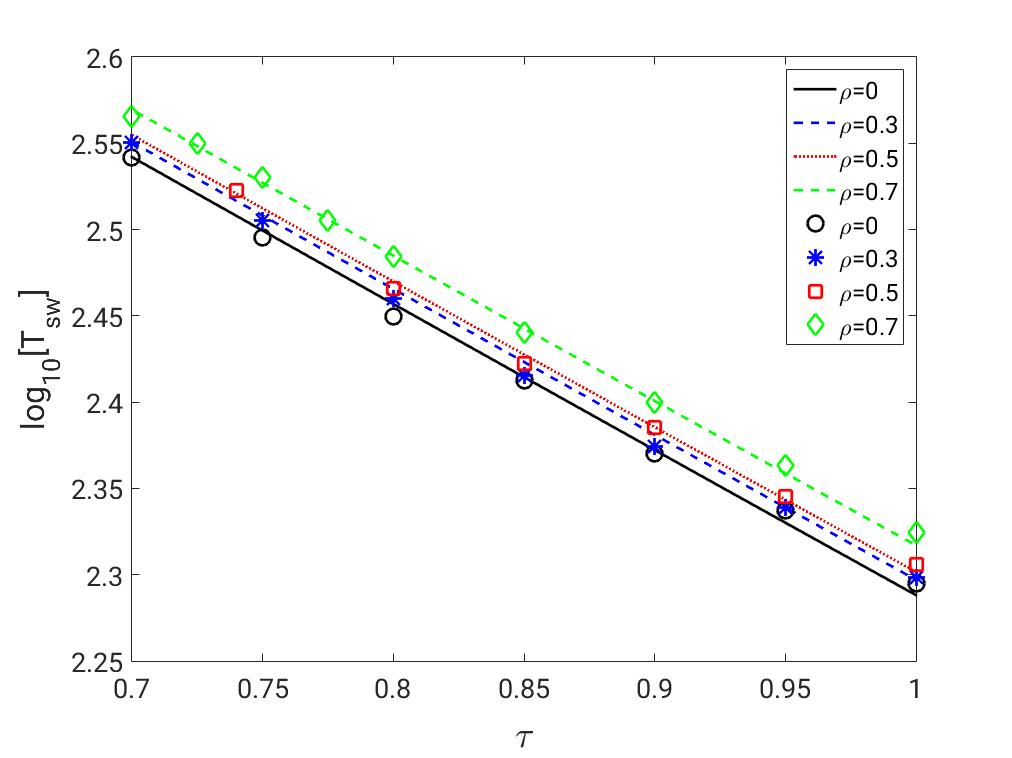}
\caption{A comparison of the first order perturbation theory in time delay $\tau$ given by (\ref{eq222}) and the $\log_{10}$ of the mean switching times obtained by numerical simulations for the uniformly distributed kernel for several values of the distribution width $\rho$. Solid lines are theoretical predictions, and symbols are numerical simulations. Other parameters are: $\gamma=0.4$, $\sqrt{2D}=0.12$, $\tau=0.8$.}\label{fig1}
\end{figure}

In this section we perform numerical simulations of the system (\ref{eq1}) with uniform and two-peak distribution kernels, given by (\ref{eq222}) in order to compare theoretical predictions for switching times. Following the same methodology as in \cite{ira15}, we calculate the mean switching times $T_{st}$ as the noise $\eta(t)$ takes the system out of the basin of attraction of the steady state $x_{A}$ to $x<x_{S}$. Numerical simulations are run for some time after the trajectory passes $x_{S}$ to ensure that the probability of its return to the basin of attraction of the stable steady state $x_{A}$ is exponentially small. Numerical simulations are performed using the Heun or stochastic predictor-corrector method \cite{Kloed,Iacus}, in which the solution to equation (\ref{eq1}) is approximated using the following iterative scheme 
\[
x_{n+1}=x_{n}+\frac{1}{2}(K_{n}+K_{n+1})h+\sqrt{2Dh}u,
\]
with a step size $h$, $u$ is $\mathcal{N}(0,1)$, and
\[
K_{n}=x_n(1-x_n)-Y_n,\quad K_{n+1}=\bar{x}_{n+1}(1-\bar{x}_{n+1})-Y_{n+1},
\]
where the approximation to $\bar{x}_{n+1}$ is given by
\[
\bar{x}_{n+1}\approx x_n+K_n h+\sqrt{2Dh}u.
\]
The terms $Y_n$ and $Y_{n+1}$ representing the delay distribution were computed using a composite trapezoidal rule in the case of uniform distribution, and for the two-peak distribution these were given by
\[
Y_n=\frac{1}{2}(x_{n-k_1}+x_{n-k_2}),\quad Y_{n+1}=\frac{1}{2}(x_{n+1-k_1}+x_{n+1-k_2}),
\]
where $k_1h=\tau-\rho$ and $k_2h=\tau+\rho$.

\begin{figure}[ht]
\includegraphics[scale=0.3]{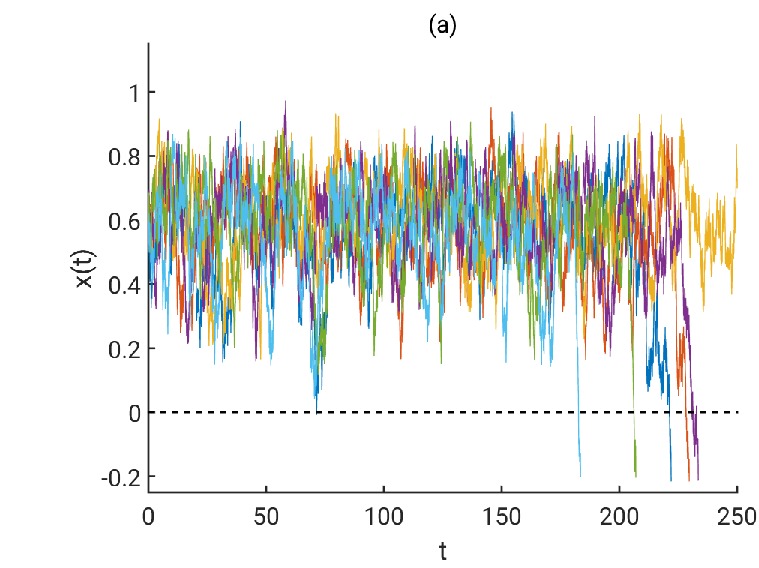}  \includegraphics[scale=0.6]{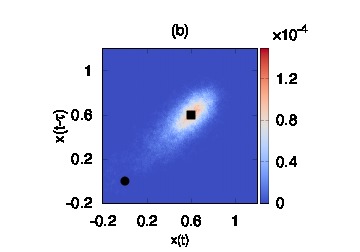} 
\caption{a) Sample switching time series in two-peak delay distribution and (b) a
  probability density of switching obtained by numerical simulations. Black squares (circles) denote the
  attractor (saddle point). Parameters are: $\gamma=0.4$, $\sqrt{2D}=0.12$, $\tau=0.8$, $\rho=0.5$.}
  \label{fig:TS_twopeak}
  \end{figure}

Examples of switching time series for several runs are shown in
Figs.~\ref{fig:TS_Uniform}(a) and \ref{fig:TS_twopeak}(a) for the uniform
and bi-modal delay distributions, respectively. One can see that the
dynamics fluctuates around a mean value of the attractor for long period of
time until at some point there is a large change in which the dynamics passes
through the saddle at the origin. The accompanying probability densities in
panel (b)
illustrate the fact that most of the time the dynamics are in the neighborhood
of the attractor. Although the probability density plots show many paths to
going from $x_A$ to the saddle at the origin, they also contain the most probable path to switch.

The switching rate is proportional to the probability of large fluctuations $\mathcal{P}_{x}[x]$ introduced in Section~\ref{sec2}, and the switching time is the inverse of the switching rate. Using the equation (\ref{rate1}), the switching time can be found as 
\[
T_{sw}=c_{sw}\exp(R_{sw}/D),
\]
where $R_{sw}$ is given by the approximation (\ref{approx22}) for the
uniformly distributed delay kernel, and the expression (\ref{approx}) in the
case of the two delay kernel approximation. Since the noise is Gaussian, the
constant $c_{sw}$ can be found using the Kramer's theory \cite{kramer} at zero
delay. This 
constant for non-zero delay corresponds to the approximate vertical shift of
the theoretical results when $T_{sw}$ is plotted on the logarithmic scale, and
is assumed to have weak dependence on noise and delay. 

\begin{figure}[ht]
\hspace{1cm}\includegraphics[width=130mm]{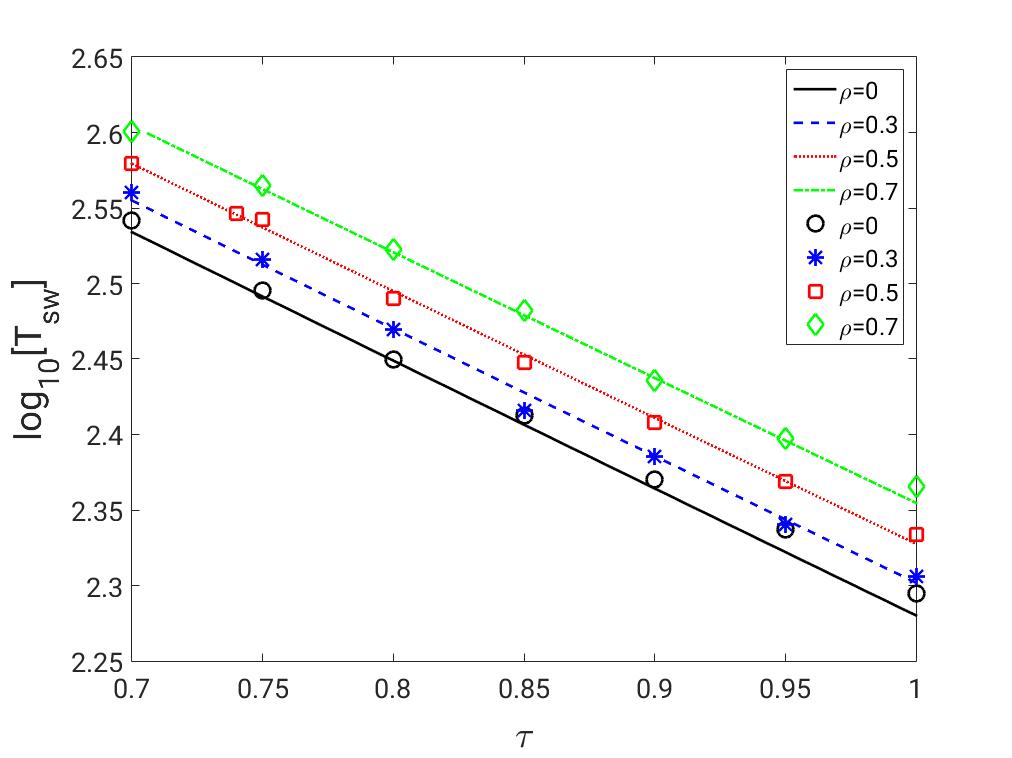}
\caption{A comparison of the first order perturbation theory in time delay $\tau$ 
  and the $\log_{10}$ of the mean switching times obtained by numerical simulations for two-peak distribution kernel for several values of the distribution width $\rho$. Solid lines are theoretical predictions, and symbols are numerical simulations. Other parameters are: $\gamma=0.4$, $\sqrt{2D}=0.12$, $\tau=0.8$.}\label{fig2}
\end{figure}

In Figs.~\ref{fig1} and \ref{fig2}, the lines represent theoretical approximations as the first order in $\tau$ perturbation theory given by the expressions (\ref{approx22}) and (\ref{approx}), and circles are the mean values of the numerical simulations taken over 2000 simulations. In Fig~\ref{fig1}, we show the comparison between the theoretical results and numerical simulations of the switching time as the function of mean time delay $\tau$ for the case of uniformly distributed kernel for several values of the distribution width $\rho$. It can be seen that increasing the mean time delay $\tau$ for any distribution width $\rho$ leads to the decrease in the switching times $T_{sw}$, and theoretical predictions align well with numerical simulations all the way up to $\tau=1$.  It is also clear  that increasing the distribution width leads to the increase in the switching times even for small values of the mean time delay $\tau$, and this increase in switching times in maintained for a whole range of admissible values of $\tau$ ($\tau<\rho$). 

A similar situation is observed in Fig~\ref{fig2}, which shows the comparison between theory and numerical simulations for the case of the two-peak distributed kernel, given by two discrete time delays separated by a time interval of $2\rho$. Again, analytical results and numerical simulations are in an excellent agreement for mean time delays up to $\tau=1$. The fastest switching times are achieved when the width $\rho$ is equal to zero, corresponding to the case of a single discrete time delay in model (\ref{eq1}), and when the mean time delay $\tau$ is large enough. As the distribution width $\rho$ becomes larger, the switching time for the trajectory to be pushed out of the basin of attraction of the steady state increases over the whole range of time delays $\tau$. This clearly shows that in delayed systems, not only the value of the mean time delay plays a role in determining the escape rates/times, but the width of the distribution also has an important influence on the overall dynamics, and the escape rates in particular. Moreover, comparing Figs.~\ref{fig1} and \ref{fig2}, it is worth noting, that while in both cases the switching times grow for larger distribution widths, the growth is more pronounced in the case of the two-peak distribution. 

\section{Discussion}

In this paper we have considered the problem of finding escape rates from the
basin of attraction of the stable steady state in a dynamical system with
distributed time delays and Gaussian noise. In the absence of noise, the
system has stable and unstable steady states, and the solutions initially  in the
basin of attraction of the stable steady state remain in the basin. When the noise is included,
most of the time the solution trajectories will fluctuate around the stable
steady state.  However, 
there exist rare instances where noise acts as a coherent force, resulting in
a  trajectory
that takes the system out of the attractor's basin of attraction. Here we have aimed to understand how particular delay
distributions influence the escape rates/times as compared to a single
discrete time delay situation. In order to calculate the escape rate, the
problem has been formulated for a general distribution kernel as a variational
problem along the optimal path, which maximizes the probability of
escape. Since  noise-induced switching occurrences are rare events, the optimal solution
to the variational problem is valid in the tail of the distribution, which is
exponential. We also note that even small changes in the action lead to
exponential changes in switching probability and switching times.

In order to make further analytical progress, having obtained the variational
equations for the extreme trajectories for a general distribution, we have
then considered two particular cases of the distribution kernel, namely,
uniform and two time delay (two-peak) distributions. For both of those
distributions, we have been able to obtain closed form analytical expressions
for the escape rates, which explicitly depend on the mean delay and
distribution width. The exemplified results hold near a trans-critical
  bifurcation point. It is known that the exponent of the  probability of escape from the attractor scales
  linearly with distance from the bifurcation point when the delay is zero. In
  contrast, we note that the scaling exponent of a saddle-node bifurcation scales as
  $\chi^{3/2}$, where $\chi$ is the distance to the bifurcation point. \cite{ira15} 

We have performed numerical simulations for both distribution kernels and
compared the results of the theoretical predictions for the escape time to the
numerically calculates ones. We have found very good agreement between theoretical prediction based on the small-delay approximation,  and numerical simulations for mean time delay values up to $\tau=1$. We have also compared theoretical and numerical results for larger values of time delay $\tau$, but found that they start to diverge for larger time delays, suggesting that the linear approximation does not work well for large time delays. 

We have found that the fastest switching times for all values of the time
delay are in the case when the distribution width is equal to zero for both
examples of the distribution kernel considered. As the distribution width is
increased, the switching times are decreasing even for the large values of the
mean time delay. Moreover, comparing the change in switching times for uniform
and two-peak distribution, we have found that while both do increase escape
times, the latter has a more profound effect. This strongly indicates that not
only the mean time delay plays a role in the switching dynamics, but the width
and the choice of the distribution are another two important factors that
influence exponentially the switching rates in time-delayed systems.

In this paper we have concentrated on the influence of the delay distribution on the switching rates in the presence of Gaussian noise and have used two particular distribution kernels. The next step would be to analyze the influence of the Gamma distributed kernel (for weak and strong Gamma kernels) and compare the results to the cases of a single discrete time delay studied in \cite{ira15}, and of the uniform and two-peak distributions considered in this paper. The model studied in this paper has a saddle point and only a single stable steady state. Another important future direction of this research would be to analyze the influence of the delay distribution on the switching times in bistable systems.
\section{Acknowledgments}
The authors gratefully  acknowledge support from the Office of Naval Research
under contract numbers  N0001418WX01225 and  N0001412WX20083, and support of the NRL Base Research Program N0001412WX30002.


%

\end{document}